\documentstyle[12pt]{article}

\begin{document}

\def\lsim{\ \matrix{<\cr\noalign{\vskip-7pt}\sim\cr} \ }
\def\gsim{\ \matrix{>\cr\noalign{\vskip-7pt}\sim\cr} \ }
\newcommand{\nc}{\newcommand}
\nc{\p}{\phi\left({\bf x},t\right)}
\nc{\ptilde}{\tilde{\phi}\left({\bf q},t\right)}
\nc{\ch}{\left|\chi({\bf q})\right|}
\nc{\lgh}{\langle\langle}
\nc{\rg}{\rangle\rangle}

\begin{titlepage}

\hfill{DFPD/TH/95-39}

\hfill{SISSA-AP/95-74}

\vskip 1cm

\centerline{\Large \bf A stochastic approach to thermal fluctuations}
\vskip 5mm
\centerline{\Large \bf during a first order electroweak  phase transition}

\vskip 1cm

\centerline{{\bf Fabrizio Illuminati $^{(a)(c)}$}\footnote{email:
illuminati@mvxpd5.pd.infn.it}
{\bf and Antonio Riotto $^{(b)(c)}$}\footnote{email:
riotto@tsmi19.sissa.it. Address after November 95:
Theoretical Astrophysics Group, NASA/Fermilab, Batavia, IL60510, USA.}}

\vskip 0.5cm

\noindent
\centerline{{\it (a) Dipartimento di Fisica, Universit\`{a}
di Padova, via F. Marzolo 8, I-35131 Padova, Italy}}

\vskip 2 mm
\noindent
\centerline{{\it (b) International School for Advanced Studies,
SISSA, via Beirut 2, I-34014 Trieste, Italy}}

\vskip 2mm
\noindent
\centerline{{\it (c) Istituto Nazionale di Fisica Nucleare, Sezione di Padova,
I-35131 Padova, Italy.}}
\vskip 1cm

\centerline{\large\bf Abstract}
\vskip 1cm
\noindent
\baselineskip 12pt
We investigate the role played by subcritical bubbles at the onset of
the electroweak phase transition. Treating the configuration modelling
the thermal fluctuations around the homogeneous zero configuration of
the Higgs field as a stochastic variable, we describe its dynamics by a
phenomenological Langevin equation. This approach allows to properly
take into account both the effects of the thermal bath on the system:
a systematic dyssipative force, which tends to erase out any initial
subcritical configuration, and  a random stochastic force responsible
for the fluctuations. We show that
the contribution to the variance $\lgh\phi^2(t)\rg_V$
in a given volume $V$
from any initial subcritical configuration is
quickly damped away and that, in the limit of long times,
$\lgh\phi^2(t)\rg_V$ approaches its equilibrium value provided by the
stochastic force and independent from the viscosity coefficient,
as predicted by the fluctuation-dissipation theorem.  In agreement
with some recent claims, we conclude that thermal fluctuations do not
affect the nucleation of critical bubbles at the onset of the
electroweak phase transition making electroweak baryogenesis scenarios
still a viable possibility to explain the primordial baryon asymmetry in
the Universe.

\end{titlepage}
\topmargin 0cm
\textwidth 154mm
\textheight 240mm
\voffset -1cm
\topskip 0mm
\baselineskip 20pt

\nc{\renc}{\renewcommand}

%
%
\nc{\etal}{\mbox{\it et al. }}
\nc{\ie}{{\it i.e.}}
\nc{\eg}{{\it e.g.}}

\renc{\thefootnote}{\arabic{footnote}}
\nc{\capt}[1]{{\bf Figure.} {\small\sl #1}}


\nc{\eqs}[2]{\mbox{Eqs.~(\ref{#1},\,\ref{#2})}}
\nc{\eq}[1]{\mbox{Eq.~(\ref{#1})}}

\nc{\figs}[2]{\mbox{Figs.~(\ref{#1},\,\ref{#2})}}
\nc{\fig}[1]{\mbox{Fig~.(\ref{#1})}}

\def\jump{\vskip 1truecm}

\nc{\tag}[1]{\label{#1} \marginpar{{\footnotesize #1}}}
\nc{\mtag}[1]{\label{#1} \mbox{\marginpar{{\footnotesize #1}}}}
\renc{\baselinestretch}{1.2}
\jot 1ex
\newlength{\overeqskip}
\newlength{\undereqskip}
\setlength{\overeqskip}{\jot}
\setlength{\undereqskip}{\jot}
%
\nc{\be}[1]{\begin{equation} \mbox{$\label{#1}$}}
\nc{\bea}[1]{\begin{eqnarray} \mbox{$\label{#1}$}}
\nc{\Section}[2]{\section{#2}\label{#1}}
\nc{\Bibitem}[1]{\bibitem{#1}}
\nc{\Label}[1]{\label{#1}}

\nc{\eea}{\vspace{\undereqskip}\end{eqnarray}}
\nc{\ee}{\vspace{\undereqskip}\end{equation}}
\nc{\bdm}{\begin{displaymath}}
\nc{\edm}{\end{displaymath}}
\nc{\dpsty}{\displaystyle}
\nc{\bc}{\begin{center}}
\nc{\ec}{\end{center}}
\nc{\ba}{\begin{array}}
\nc{\ea}{\end{array}}
\nc{\bab}{\begin{abstract}}
\nc{\eab}{\end{abstract}}
\nc{\btab}{\begin{tabular}}
\nc{\etab}{\end{tabular}}
\nc{\bit}{\begin{itemize}}
\nc{\eit}{\end{itemize}}
\nc{\ben}{\begin{enumerate}}
\nc{\een}{\end{enumerate}}
\nc{\bfig}{\begin{figure}}
\nc{\efig}{\end{figure}}
%
%
\nc{\arreq}{&\!=\!&}
\nc{\arrmi}{&\!-\!&}
\nc{\arrpl}{&\!+\!&}
\nc{\arrap}{&\!\!\!\approx\!\!\!&}
\nc{\non}{\nonumber\\*}
\nc{\align}{\!\!\!\!\!\!\!\!&&}

\def\lsim{\; \raise0.3ex\hbox{$<$\kern-0.75em
      \raise-1.1ex\hbox{$\sim$}}\; }
\def\gsim{\; \raise0.3ex\hbox{$>$\kern-0.75em
      \raise-1.1ex\hbox{$\sim$}}\; }
\nc{\DOT}{\hspace{-0.08in}{\bf .}\hspace{0.1in}}
\nc{\Laada}{\hbox {$\sqcap$ \kern -1em $\sqcup$}}
\nc\loota{{\scriptstyle\sqcap\kern-0.55em\hbox{$\scriptstyle\sqcup$}}}
\nc\Loota{{\sqcap\kern-0.65em\hbox{$\sqcup$}}}
\nc\laada{\Loota}
\nc{\qed}{\hskip 3em \hbox{\BOX} \vskip 2ex}
\def\Re{{\rm Re}\hskip2pt}
\def\Im{{\rm Im}\hskip2pt}
\nc{\real}{{\rm I \! R}}
\nc{\Z}{{\sf Z \!\!\! Z}}
\nc{\complex}{{\rm C\!\!\! {\sf I}\,\,}}
\def\bigid{\leavevmode\hbox{\small1\kern-3.8pt\normalsize1}}
\def\id{\leavevmode\hbox{\small1\kern-3.3pt\normalsize1}}
\nc{\slask}{\!\!\!/}
\nc{\bis}{{\prime\prime}}
\nc{\pa}{\partial}
\nc{\na}{\nabla}
\nc{\ra}{\rangle}
\nc{\la}{\langle}
\nc{\goto}{\rightarrow}
\nc{\swap}{\leftrightarrow}

\nc{\EE}[1]{ \mbox{$\cdot10^{#1}$} }
\nc{\abs}[1]{\left|#1\right|}
\nc{\at}[2]{\left.#1\right|_{#2}}
\nc{\norm}[1]{\|#1\|}
\nc{\abscut}[2]{\Abs{#1}_{\scriptscriptstyle#2}}
\nc{\vek}[1]{{\rm\bf #1}}
\nc{\integral}[2]{\int\limits_{#1}^{#2}}
\nc{\inv}[1]{\frac{1}{#1}}
\nc{\dd}[2]{{{\partial #1}\over{\partial #2}}}
\nc{\ddd}[2]{{{{\partial}^2 #1}\over{\partial {#2}^2}}}
\nc{\dddd}[3]{{{{\partial}^2 #1}\over
	{\partial #2 \partial #3}}}
\nc{\dder}[2]{{{d #1}\over{d #2}}}
\nc{\ddder}[2]{{{d^2 #1}\over{d {#2}^2}}}
\nc{\dddder}[3]{{d^2 #1}\over
	{d #2 d #3}}
\nc{\dx}[1]{d\,^{#1}x}
\nc{\dy}[1]{d\,^{#1}y}
\nc{\dz}[1]{d\,^{#1}z}
\nc{\dl}[1]{\frac{d\,^{#1}l}{(2\pi)^{#1}}}
\nc{\dk}[1]{\frac{d\,^{#1}k}{(2\pi)^{#1}}}
\nc{\dq}[1]{\frac{d\,^{#1}q}{(2\pi)^{#1}}}

\nc{\cc}{\mbox{$c.c.$ }}
\nc{\hc}{\mbox{$h.c.$ }}
\nc{\cf}{cf.\ }
\nc{\erfc}{{\rm erfc}}
\nc{\Tr}{{\rm Tr\,}}
\nc{\tr}{{\rm tr\,}}
\nc{\pol}{{\rm pol}}
\nc{\sign}{{\rm sign}}
\nc{\bfT}{{\bf T }}
\def\GeV{{\rm\ GeV}}
\def\MeV{{\rm\ TeV}}
\def\keV{{\rm\ keV}}
\def\TeV{{\rm\ TeV}}

\nc{\cA}{{\cal A}}
\nc{\cB}{{\cal B}}
\nc{\cD}{{\cal D}}
\nc{\cE}{{\cal E}}
\nc{\cG}{{\cal G}}
\nc{\cH}{{\cal H}}
\nc{\cL}{{\cal L}}
\nc{\cO}{{\cal O}}
\nc{\cT}{{\cal T}}
\nc{\cN}{{\cal N}}
%
\nc{\rvac}[1]{|{\cal O}#1\rangle}
\nc{\lvac}[1]{\langle{\cal O}#1|}
\nc{\rvacb}[1]{|{\cal O}_\beta #1\rangle}
\nc{\lvacb}[1]{\langle{\cal O}_\beta #1 |}
\nc{\bb}{\bar{\beta}}
\nc{\bt}{\tilde{\beta}}
\nc{\ctH}{\tilde{\cal H}}
\nc{\chH}{\hat{\cal H}}
%
\nc{\1}{\aa}
\nc{\2}{\"{a}}
\nc{\3}{\"{o}}
\nc{\4}{\AA}
\nc{\5}{\"{A}}
\nc{\6}{\"{O}}
%
\nc{\al}{\alpha}
\nc{\g}{\gamma}
\nc{\Del}{\Delta}
\nc{\e}{\epsilon}
\nc{\eps}{\epsilon}
\nc{\lam}{\lambda}
\nc{\om}{\omega}
\nc{\Om}{\Omega}
\nc{\ve}{\varepsilon}
\nc{\mn}{{\mu\nu}}
\nc{\k}{\kappa}
\nc{\vp}{\varphi}

%
%
\nc{\rf}[1]{(\ref{#1})}
\nc{\nn}{\nonumber \\*}
\nc{\bfB}{\bf{B}}
\nc{\bfv}{\bf{v}}
\nc{\bfx}{\bf{x}}
\nc{\bfy}{\bf{y}}
\nc{\vx}{\vec{x}}
\nc{\vy}{\vec{y}}
\nc{\oB}{\overline{B}}
\nc{\oI}{\overline{I}}
\nc{\oR}{\overline{R}}
\nc{\rar}{\rightarrow}
\nc{\ti}{\times}
\nc{\slsh}{\hskip-5pt/}
\nc{\sm}{Standard~Model~}
\nc{\MP}{M_{\rm Pl}}
\nc{\tp}{t_{\rm Pl}}
\nc{\ave}{\bar{E}}
\def\mes#1{{d^3{#1}\over (2\pi )^3}}
\renc{\min}{p_{\rm min}}
\renc{\max}{p_{\rm max}}
\nc{\pmin}{p_{\rm min}}
\nc{\pmax}{p_{\rm max}}
\nc{\fo}{f_0}
\nc{\foi}{f_{0,i}\,}
\nc{\fop}{f_0^P}
\nc{\fou}{f_0^U}
\def\sepand{\rule{14cm}{0pt}\and}
\nc{\eff}{{\rm eff}}
\nc{\MT}{M_{\rm T}}
\nc{\ML}{M_{\rm L}}
\nc{\kk}{\vek{k}}
\nc{\pp}{{\rm p}}
\nc{\cb}{critical bubble~}
\nc{\cbs}{critical bubbles~}
\nc{\scb}{subcritical bubble~}
\nc{\scbs}{subcritical bubbles~}
\nc{\vv}{\\}
%

\leftline{\large\bf 1. Introduction}
\vskip 5mm
\baselineskip 20pt

Nucleation of critical bubbles during a first order electroweak
phase transition
has received much attention since the discovery of the  possibility
for electroweak baryogenesis \cite{ckn}. Indeed, one of the basic
ingredients for the generation of the baryon asymmetry (apart from the
requirement of
baryon- and CP-violating interactions) is the presence of
an out-of-equilibrium state \cite{sak} which, during the
first order electroweak phase
transition with supercooling, is
attained by critical bubbles expanding in the thermal
bath of the unbroken phase.

However, less attention has been paid to
the environment where the critical bubble nucleation is supposed to occur.
Since critical bubbles have a finite size, phase transitions
are highly local phenomena. Fluctuations of the Higgs scalar
field $\phi$ around the origin $\phi=0$ with spatial correlations
comparable to the critical bubble size may be expected to be important
for bubble nucleation. Also, if thermal fluctuations are too large,
any perturbative scheme
could break down. In such a situation, the
prediction of a first order phase transition
becomes suspect, with the possibility that the entire scenario of electroweak
baryogenesis might be  invalidated.

On the other hand, while the presence of thermal fluctuations
in any thermodynamical system is
undisputed, their role in the dynamics of weakly first order phase
transitions is still controversial.

In ref. \cite{gleiser}, it was first conjectured
that  statistical fluctuations around equilibrium
may be described by spherically symmetric configurations roughly
extending over a correlation volume, where
the correlation length is given by the inverse temperature dependent
mass of the Higgs field: $\xi(T)=m^{-1}(T)$.

These fluctuations are
referred to as subcritical bubbles. Their
amplitude was estimated in ref. \cite{gleiser},
where it was concluded
that they are dominant if the Higgs mass $M_H$ is larger than $\sim 80$ GeV,
whence the  fraction which the asymmetric vacuum
occupies in the neighborhood of
the critical temperature becomes of the order of unity. Therefore it was
concluded that
critical bubbles cannot be generated due to the inhomogeneities of the
background field.

In ref. \cite{gleiser}, however, the continuous
disappearance of the subcritical
bubbles was not accounted for. This can happen in two ways: the
subcritical bubbles, being unstable configurations, tend to shrink; they
are also subject to constant thermal bombardment so that they
may disappear simply because of thermal noise.

Following general
principles and estimating the scalar field
two-point function, {\it i.e.} the variance of a gaussian fluctuation
distribution, computed in a correlation volume,
Dine {\it et al.} \cite{dine} and Anderson \cite{and} have later argued that
subcritical bubbles do not affect the nucleation of
critical bubbles in an appreciably way.

Subsequently, Gelmini and Gleiser \cite{gel} rekindled the issue
adopting a different point of view based on modelling by a set of
Boltzmann equations the evolution with time $t$
of the number density
$n(R,t)$ of subcritical bubbles with a certain radius $R$.

Under a specific assumption about the form of the destruction
rate due to thermal noise, they
found that for Higgs masses below $\sim$ 55 GeV
the approach to equilibrium is dominated by shrinking. Unfortunately,
for the interesting range of Higgs masses dictated by the experimental
constraints coming from LEP, $M_H> 60$ GeV \cite{lep}, their analysis is
inconclusive since the approximations adopted break down.

Recently, Enqvist {\it et al.} \cite{noi} re-estimated
the amplitude, average size and formation rate of subcritical bubbles
taking into account the crucial role played by thermalization. Their
starting point was the
the observation that at the microscopic level,
the true origin of the dominant thermal fluctuation is the perpetual
creation and annihilation of spherical subcritical bubbles. Thus one
should identify the typical amplitude and size of these bubbles
with the values estimated
by a statistical ensemble averaging, instead of assuming, for
instance, the size fixed and equal to the
correlation length $\xi(T)$.

They also observed that
a large subcritical bubble
should resemble the critical one around the critical temperature:
when $R$ increases, the
form of the subcritical bubble should deform smoothly to reproduce
the critical configuration when
$R=R_c$, $R_c$ being the critical radius. Thus, it was
found that the  average size $\langle R\rangle$  of
the subcritical bubbles is much larger than the correlation length,
and that the average amplitude $\langle \phi_0^2\rangle^{1/2}$ at the core
is much smaller than any previous estimate.

These results led the authors of ref.
\cite{noi} to conclude that  thermal fluctuations do not hinder the
electroweak first order phase transition. Of course, treating the size
and the amplitude of subcritical bubbles as statistical degrees
of freedom of an {\it equilibrium} ensamble at a fixed temperature $T$
does not allow to follow the time evolution of such degrees
of freedom. Thus, since in the approach of ref. \cite{noi} there
is no dynamics involved, $\langle R\rangle$ and $\langle \phi_0^2\rangle^{1/2}$
should be interpretated as the most probable initial conditions
for any configuration
describing a subcritical bubble generated at the time $t=0$.

Analogous conclusions to those of ref. \cite{noi} have been very
recently
obtained by Bettencourt \cite{bet}, who, along the same line followed
by Hindmarsh and Rivers \cite{hr} for the $\lambda\phi^4$
theory, computed the probability
for fluctuations of the Standard Model
Higgs field, averaged over a given spatial
scale, to exceed a specified value. He found that the
probability for the Higgs scalar field
to fluctuate from the symmetric to the asymmetric minimum before
the latter becomes stable is very small for Higgs masses of order
of those of the $W^{\pm}$ and $Z^0$ bosons, whereas the converse is more
likely.

The aim of the present paper is to take a further step
in the investigation of
the role played by subcritical bubbles during the onset of the
electroweak phase transition. In particular, we wish to present
an analytic treatment of the dynamics of subcritical bubbles, described
by a suitably coarse-grained configuration $\p$, in a random
environment.

Indeed, thermal fluctuations are the manifestation of the
interaction between the system and the surrounding environment. The
peculiarity of the thermal bath is that a virtually infinite number of
degrees of freedom takes part in the exchanges of energy with the
system. A familiar example \cite{kardar}
are the molecules of a gas or a fluid in
which the system (the brownian particle) is embedded. It is well known
that the impacts between the the system and the surrounding particles
cause two effects: a systematic dissipative force (friction) and a
random force, responsible for the fluctuations.

To describe the dynamics of subcritical fluctuations in a thermal bath
around the homogeneous
configuration $\phi=0$, we will then
assume a classical Markovian Langevin equation
\begin{equation}
\ddot{\phi}({\bf x},t)-
\nabla^2\p+\eta\dot{\phi}({\bf x},t)=-\frac{\partial V(\phi,T)}{\partial
\phi}+\gamma\left({\bf x},t\right),
\label{langevin}
\end{equation}
where $V(\phi,T)$ is the potential associated to the
the scalar field $\p$ and
$\eta$ is
the viscosity coefficient which takes into account
the dissipative effects.

The fluctuations are modelled by introducing a stochastic
additive force (noise term)
$\gamma\left({\bf x},t\right)$ characterized by a gaussian
distribution with
\begin{eqnarray}
\langle\gamma\left({\bf x},t\right)\rangle&=&0,\nonumber\\
\langle\gamma\left({\bf x},t\right)\gamma\left({\bf x}^{\prime},t
^{\prime}\right)\rangle&=&2\:D\:\delta^3({\bf x}-{\bf x}^{\prime})
\:\delta(t-t^{\prime}),
\label{correlation}
\end{eqnarray}
where $D$ is the diffusion coefficient.

The Langevin equation (1) describing the dynamics of
a subcritical bubble in contact with a thermal bath is similar to
the Kardar-Parisi-Zhang equation \cite{kpz} modelling the evolution of
fluctuations around the average profile
of a growing interface, which, in our case, is
represented by the homogeneous configuration $\phi=0$.

Since the dynamics is expected to bring the system
into equilibrium with the thermal bath at long times, the Langevin
equation describes correctly the long-time dynamics and reproduce the
correct equilibrium behaviour if the fluctuation-dissipation condition
$D=\eta\:T$, expressing the common origin of dissipation and
fluctuation, is satisfied \cite{hoh}.

Equation (1)
naturally
incorporates different crucial informations: {\it i)}
subcritical bubbles are subject to constant thermal
bombardment so that they can be rapidly thermalized and disappear with a
relaxation
rate $\eta$ (this was also accounted for in ref. \cite{noi} where,
however, no dynamics was involved); {\it ii)} subcritical bubbles, being
unstable configurations, tend to shrink; {\it iii)} the crucial role
played by the noise term  $\gamma\left({\bf x},t\right)$ responsible for
fluctuations. It tends to
contrast the damping term
$\eta\dot{\phi}$ and determines the form of the
fluctuation $\langle \phi^2({\bf x},t) \rangle$ at long
times as well as its final equilibrium value.

A few comments are in order here about our choice of eq. (1). In writing
it we have assumed that the system is Markovian:
the correlation time scale for the noise is smaller than the typical
relaxation time scale for the system. We have also assumed the noise to
be additive.

Recent works indicates that one should expect departures from the
Langevin equation written above \cite{vari}, although the details are
very sensitive to the model one starts with. For istance, in a
$\lambda\phi^4$ theory, the Langevin equation describing
the dynamics of the long wavelength modes evolving in the bath
formed by the short wavelength modes is characterized by a colored
and multiplicative noise
and by a space-time dependent viscosity coefficient (see Gleiser and Ramos
in \cite{vari}).

In this paper we will adopt eq. (1) as a first step. In fact, we do not
expect that the nature of the thermal noise will change the final
equilibrium properties of the system, even if the relaxation
time-scales can be changed. Since the physical results will be related
to the final equilibrium value of $\langle \phi^2({\bf x},t) \rangle$,
we believe that they will be not
affected by more complicated (even if more physical)
representations of the coupling of the field $\p$ to the thermal bath.

This point of view is also motivated by the independence of
the statistical moments of the $\p$ field at long times
from the viscosity coefficient $\eta$,
as implicit in the fluctuation-dissipation theorem.

For sake of clarity we have decided to address the issue of dynamics of
subcritical bubbles in the electroweak theory after having described
the same issue for simpler theories.

The paper is then organized as follows. In Subsection 2.1 we  describe in
details, for a free scalar field theory in the absence of thermal
environments, the dynamics of a configuration representing an initial
departure from the minimum energy equilibrium configuration.
This allows us to evaluate the typical shrinking lifetime
$\tau_{sh}$ of such a
configuration.

In Subsection 2.2 we let the free scalar field be coupled
to a thermal environment and follow
the dynamics of the fluctuation at short and long times.
We determine the asymptotic value of $\langle \phi^2({\bf x},t) \rangle$
 in a
given volume
for $t\gg 1/\eta$, showing explicitly that it is independent of time
and equal to its equilibrium value.

In Section 3 we make use of the results obtained in Section 2
to describe, through a
self-consistent Hartree approximation, the strength of the thermal
fluctuations for the $\lambda\phi^4$ theory.

In Section 4 we present our
results for the electroweak theory, from which one can conclude that
subcritical bubbles do not affect nucleations of critical bubbles in a
appreciably way.

We finally draw in Section 5
our conclusions and perspectives
for future work.

\vskip 1cm
\leftline{\large\bf 2. Free scalar field: linear dynamics}
\vskip 0.4cm
\leftline{\bf 2.1 Free evolution at zero temperature}
\vskip 5mm
In this Section we want to analyze the evolution of a
spherically-symmetric configuration in the linear regime and
absence of coupling to the thermal bath. We already
know from Derrick's theorem \cite{der}, which forbids in four dimensions
the existence of static, finite-energy configurations
for models containing a simple real
scalar field, that such configurations must be unstable and decay with
a relaxation time $\tau_{sh}$, that we now
want to estimate \cite{comment}.

Given the quadratic mass term ($m^2>0$)
\begin{equation}
\label{linear}
V(\phi)=\frac{m^2}{2}\phi^2,
\end{equation}
the associated Klein-Gordon equation is given by
\begin{equation}
\label{kg}
\ddot{\phi}({\bf x},t)-
\nabla^2\p+m^2\p=0.
\end{equation}
We model the initial deviation at $t=0$ from the homogeneous
minimum-energy configuration $\phi=0$ by a gaussian shape of the
type
\begin{eqnarray}
\label{initial}
\phi({\bf x},0)&=&\phi_0\:{\rm e}^{-\left|{\bf
x}\right|^2/2\:R_0^2},\nonumber\\
\dot{\phi}({\bf x},0)&=&0,
\end{eqnarray}
{\it i.e.} we imagine a deviation formed initially at rest
with initial radius $R_0$ and
amplitude $\phi_0$ at its core.
Eq. (\ref{kg}) is most easily solved by examining the Fourier components
\begin{equation}
\label{fourier}
\ptilde=\int\:d^3{\bf x}\:{\rm e}^{i{\bf q}\cdot{\bf x}}\:\p,
\end{equation}
which evolve according to
\begin{equation}
\label{equationfourier}
\ddot{\tilde{\phi}}\left({\bf q},t\right)+\left(\left|{\bf
q}\right|^2+m^2\right)\ptilde=0.
\end{equation}
The solution of equation (\ref{equationfourier}) is easily found to be
\begin{equation}
\label{solutionkg}
\ptilde=A({\bf q})\cos\left(\sqrt{\left|{\bf
q}\right|^2+m^2}\:t\right)+B({\bf q})\sin\left(\sqrt{\left|{\bf
q}\right|^2+m^2}\:t\right).
\end{equation}
The initial conditions described in eq. (\ref{initial}) fix
$A({\bf q})$ and $B({\bf q})$
\begin{eqnarray}
\label{ab}
A({\bf q})&=&\left(2\pi\right)^{3/2}   \phi_0\:R_0^3\:
{\rm e}^{-\left|{\bf
q}\right|^2R_0^2/2},\nonumber\\
B({\bf q})&=&0,
\end{eqnarray}
which, when plugged into eq. (\ref{solutionkg}) allow us to
investigate the behaviour of the bubble's core with time
\begin{equation}
\phi({\bf 0},t)=\sqrt{\frac{2}{\pi}}\:\phi_0 R_0^3\: \int_0^{\infty}
d\left|{\bf q}\right|\:\left|{\bf q}\right|^2\:{\rm e}^{-\left|{\bf
q}\right|^2R_0^2/2}\cos\left(\sqrt{\left|{\bf
q}\right|^2+m^2}\:t\right).
\end{equation}
Since the integral is dominated by small values of $\left|{\bf
q}\right|$, $\left|{\bf q}\right|\lsim\sqrt{2}R_0^{-1}$, we can
approximate the argument of $\cos\left(\sqrt{\left|{\bf
q}\right|^2+m^2}\:t\right)$ for $R_0\gsim\sqrt{2}m^{-1}$ and write
\begin{eqnarray}
\phi({\bf 0},t)&\simeq&\sqrt{\frac{2}{\pi}}\:\phi_0 R_0^3\:
\int_0^{\infty}
d\left|{\bf q}\right|\:\left|{\bf q}\right|^2\:
{\rm Re}\left({\rm e}^{i m t}\:{\rm e}^{-\left|{\bf
q}\right|^2R_0^2/2}\:{\rm e}^{i \left|{\bf q}\right|^2
t/2m}\right)\nonumber\\
&& \nonumber \\
&=&\phi_0 R_0^3\:\frac{\cos\left[mt+\frac{3}{2}\:
{\rm tan}^{-1}\left(t/mR_0^2\right)\right]}{\left(
R_0^4+\frac{t^2}{m^2}\right)^{3/4}}.
\end{eqnarray}
Thus, given an initial deviation
from the stable configuration $\phi=0$
with sufficiently large initial size $R_0$
and in absence of coupling to a thermal
bath, the amplitude at the core
decays as $t^{-3/2}$ and the unstable
configuration shrinks by radiating
away its initial energy after a  lifetime \cite{comment}
\begin{equation}
\tau_{sh}\sim m\: R_0^2\gsim \frac{1}{m}.
\end{equation}
\vskip 1cm
\leftline{\bf 2.2 Coupling to the thermal bath}
\vskip 5mm
The next step is to investigate how the dynamics of the
spherically-symmetric fluctuations
around the homogeneous configuration $\phi=0$ changes when
the system is coupled to a thermal bath.

The Langevin equation (1) reads
\begin{equation}
\ddot{\phi}({\bf x},t)-
\nabla^2\p+\eta\dot{\phi}({\bf x},t)+m^2\p=\gamma\left({\bf x},t\right).
\label{lanlin}
\end{equation}
One can imagine the above equation to emerge from an effective theory
where the fields (for instance, scalars different
from $\p$ or fermions) which $\p$ is coupled to have
been integrated out leaving an effective potential
approximated at high temperature
by eq. (\ref{linear}). In such a case $m^2$ has to be
understood as a function
of the temperature, $m^2(T)\sim g^2 T^2$, $g$ being the generic coupling
constant of $\p$ to the other fields of the underlying
original theory.

We wish to point out once more
that the form of the Langevin equation arising from an
effective theory is strongly model-dependent.
Nevertheless, since we are interested
in the equilibrium value of the fluctuation
$\langle \phi^2({\bf x},t) \rangle$, we are confident that our result
will not be affected by using eq. (\ref{lanlin}) to describe
the dynamics of subcritical bubbles.

Again, eq. (\ref{lanlin}) is most easily solved by examining the
Fourier components of the field $\p$, see eq. (\ref{fourier}).
If we define
\begin{equation}
\label{fouriernoise}
\tilde{\gamma}({\bf q},t)=\int\:d^3{\bf x}\:{\rm e}^{i{\bf q}\cdot{\bf x}}\:
\gamma({\bf x},t),
\end{equation}
the Fourier transformed noise, the equation to solve is
\begin{equation}
\label{master}
\ddot{\tilde{\phi}}\left({\bf q},t\right)+
\eta\dot{\tilde{\phi}}\left({\bf q},t\right)+
\left(\left|{\bf
q}\right|^2+m^2\right)\ptilde=\tilde{\gamma}({\bf q},t).
\end{equation}
We define the function $\chi$ as
\begin{equation}
\label{chi}
\chi({\bf q})\equiv \eta^2-4\left(\left|{\bf q}\right|^2+m^2\right),
\end{equation}
and assume
\begin{equation}
\label{condition}
\eta<2\:m.
\end{equation}
The case $\eta>2\:m$, though straightforward,
is computationally more
involved. On the other hand, it is physically the least
interesting case, since any oscillatory behaviour is
quickly dominated by the exponentially damped factors.
Moreover, it obviously leads to the same results as far as the
final equilibrium values are concerned.

The solution of eq. (\ref{master}) can be cast into the form
\begin{equation}
\label{solution}
\ptilde=\tilde{\phi}_{ho}({\bf q},t)+\tilde{\phi}_\gamma({\bf q},t),
\end{equation}
where
\begin{eqnarray}
\label{three}
\tilde{\phi}_{ho}({\bf q},t)&=& A({\bf q})\:{\rm e}^{-\eta t/2}\:
\cos\frac{\sqrt{\ch}}{2}t \: + \: B({\bf q})\:{\rm e}^{-\eta t/2}\:
\sin\frac{\sqrt{\ch}}{2}t,\nonumber\\
&& \nonumber \\
\tilde{\phi}_\gamma({\bf q},t)&=& -2\:{\rm e}^{-\eta t/2}\:
\cos\frac{\sqrt{\ch}}{2}t\:\int_0^t d\tau\:\frac{{\rm e}^{\eta
\tau/2}}
{\sqrt{\ch}}\:\sin\frac{\sqrt{\ch}}{2}\tau\:\gamma({\bf
q},\tau) \nonumber \\
&& \nonumber \\
&+&2\:{\rm e}^{-\eta t/2}\:
\sin\frac{\sqrt{\ch}}{2}t\:\int_0^t d\tau\:\frac{{\rm e}^{\eta
\tau/2}}
{\sqrt{\ch}}\:\cos\frac{\sqrt{\ch}}{2}\tau\:\gamma({\bf
q},\tau)\nonumber.
\end{eqnarray}
\begin{equation}
\end{equation}
Note that $\tilde{\phi}_\gamma({\bf q},t)$ does not depend upon
the initial conditions described by eq.
(\ref{initial}), while they are embodied in
$\tilde{\phi}_{ho}({\bf q},t)$ through the amplitudes
$A({\bf q})$ (given by eq. (\ref{ab})) and
\begin{equation}
\label{b}
B({\bf q})=\frac{\eta}{\sqrt{\ch}}\:A({\bf q}).
\end{equation}
When computing the mean values $\langle\cdot\cdot\rangle$
with respect to the gaussian measure
of the noise $\gamma({\bf
x},t)$, some care is needed.
Indeed, as noted in the Introduction, one has to assume
that the correlation time scale for the noise $\tau_{\gamma}$  is
smaller than the typical relaxation time of the
free system $\tau_{sh}\simeq R_0^2\:m$ as determined
in Subsect. 2.1.

In the case under examination,
$\eta<2\:m$, this is true only for configurations
with very large initial radius
$R_0\gg m^{-1}$; then, from now on we will assume this to be the
case\footnote{Of
course, in the opposite limit $\eta>2\:m$, the correlation time scale
for the noise is much smaller than any other time scales and no
additional condition on $R_0$ is required.}, {\it i.e.}
$\tau_{\gamma}\ll\tau_{sh}$.

{}From eq. (\ref{three}) we see that fluctuations in each mode
decay
with a relaxation time given by $\tau_{\gamma}\simeq 1/\eta$.
This implies for the following decay law for the average values:
\begin{equation}
\label{relaxation}
\langle\ptilde\rangle\sim {\rm exp}(-\eta t/2),
\end{equation}
as expected
from the presence of the damping term in the Langevin equation.
This is similar to what happens for a particle thrown into
a thermal bath with an initial velocity $\vec{v}_0$:
after a sufficiently long time, and independently from
the values of $\vec{v}_0$,
the particle behaves like a brownian system with
a configurational average
$\langle \vec{X}\rangle=0$ and a nonvanishing
fluctuation $\langle \vec{X}^2\rangle$
determined by the surrounding noise.

The two-point correlation function at equal time is given by
\begin{eqnarray}
\label{twopoint}
\langle\p \phi({\bf x}^{\prime},t)\rangle&=&
\int\:\frac{d^3{\bf q}}{(2\pi)^3}\:{\rm e}^{-i{\bf q}\cdot{\bf x}}\:
\int\:\frac{d^3{\bf q}^{\prime}}{(2\pi)^3}\:
{\rm e}^{-i{\bf q}^{\prime}\cdot{\bf x}^{\prime}}\:
\langle\ptilde \tilde{\phi}({\bf q}^{\prime},t)\rangle\nonumber \\
& & \nonumber \\
&=&\phi_{ho}({\bf x},t)\phi_{ho}({\bf x}^{\prime},t)+
\langle\p \phi({\bf x}^{\prime},t)\rangle_{\gamma},
\end{eqnarray}
where we have defined
\begin{equation}
\langle\p \phi({\bf x}^{\prime},t)\rangle_{\gamma}=
\int\:\frac{d^3{\bf q}}{(2\pi)^3}\:{\rm e}^{-i{\bf q}\cdot{\bf x}}\:
\int\:\frac{d^3{\bf q}^{\prime}}{(2\pi)^3}\:
{\rm e}^{-i{\bf q}^{\prime}\cdot{\bf x}^{\prime}}\:
\langle\tilde{\phi}_{\gamma}({\bf q},t)
\tilde{\phi}_{\gamma}({\bf q}^{\prime},t)\rangle.
\label{a}
\end{equation}
Making use of the two-point correlation function for the Fourier
transformed noise
\begin{equation}
\langle\gamma\left({\bf q},t\right)\gamma\left({\bf q}^{\prime},t
^{\prime}\right)\rangle=2\:(2\pi)^3\:
\eta\:T\:\delta^3({\bf q}+{\bf q}^{\prime})
\:\delta(t-t^{\prime}),
\label{correlationfourier}
\end{equation}
eq. (\ref{a}) becomes
\begin{eqnarray}
\langle\p \phi({\bf x}^{\prime},t)\rangle_{\gamma}&=&
4\:T\:\int\:\frac{d^3{\bf q}}{(2\pi)^3}\:{\rm e}^{-i{\bf q}\cdot
({\bf x}-{\bf x}^{\prime})}\:\frac{1}{\ch}\:
\left[\frac{\ch}{\eta^2+\ch}-{\rm e}^{-\eta t}\right.\nonumber \\
& & \nonumber \\
&+&\left.{\rm e}^{-\eta t}\:
\frac{\eta^2\:\cos\sqrt{\ch}t-
\eta\:\sqrt{\ch}\:\sin\sqrt{\ch}t}{
\eta^2+\ch}\right].\nonumber
\end{eqnarray}
\begin{equation}
\end{equation}
We are now in the position to study the evolution
with time of the magnitude of the fluctuation around the $\phi=0$ state.
A natural measure  of it is given by the two-point connected Green's
function, coarse-grained on a volume $V$, defined as
\begin{eqnarray}
\label{green}
\lgh\phi^2(t)\rg_V&=&
\frac{1}{V^2}\:
\int\:d^3{\bf x}\:\int\:d^3{\bf x}^{\prime}\:
\langle\p \phi({\bf x}^{\prime},t)\rangle\:I({\bf x})\:
I({\bf x}^{\prime})\nonumber \\
& & \nonumber \\
&=&\lgh\phi^2(t)\rg_V^{ho}
+\lgh\phi^2(t)\rg_V^{\gamma},
\end{eqnarray}
where the definition of
$\lgh\phi^2(t)\rg_V^{ho}$ and
$\lgh\phi^2(t)\rg_V^{\gamma}$
can be easily read from eq. (\ref{twopoint}) and
\begin{equation}
\label{window}
I({\bf x})=\sqrt{\frac{2}{9\pi}}\:{\rm e}^{-\left|{\bf
x}\right|^2/2\:R^2},
\end{equation}
is a window function modelling the volume $V=(4\pi/3)R^3 $
over which we test the
magnitude of the fluctuations\footnote{The coefficient $\sqrt{2/9\pi}$
in eq. (\ref{window}) is chosen to make easier
the comparison between our results in Section 4
and those in ref. \cite{bet} in the case of
the fluctuations at the onset of
the electroweak phase transition.}.

The system as a whole is taken to have a
volume much larger than $V$. If we denote by $\xi\simeq 1/m$ the correlation
length, we can parametrize the number of correlation volumes by writing
$R=\beta\xi$, where $\beta$ is a positive number. Since, at any time,
the field $\phi$ is correlated over some correlation
length, it will be sufficient
to take $\beta>1$.

{}From eq. (\ref{green}) we see
that the variance $\lgh\phi^2(t)\rg_V$ receives two contributions:
$\lgh\phi^2(t)\rg_V^{ho}$ describing the contribution coming
from the gaussian configuration formed at $t=0$ and
$\lgh\phi^2(t)\rg_V^{\gamma}$ expressing the
contribution due to the noise force. As a consequence, the latter
does not
depend on the initial conditions.

Let us now imagine that the gaussian configuration
(\ref{initial}) represents a deviation from the equilibrium
value formed at $t=0$ due to thermal fluctuations. To follow the
dynamics of $\lgh\phi^2(t)\rg_V$ we can then
identify two different time regimes:
\vskip 5mm
\begin{flushleft}
{\it i)} Short times:
\end{flushleft}
For $t\ll 1/\eta$, none of the modes of
$\lgh\phi^2(t)\rg_V^{ho}$ has yet relaxed, but $\lgh\phi^2(t)
\rg_V^{ho}$ starts to decrease as $t^{2}$, that is
\begin{eqnarray}
& { \, } & \lgh\phi^2(t)\rg_V^{ho}
-\lgh\phi^2(0)\rg_V^{ho} \simeq \nonumber \\
& & \nonumber \\
& - & t^{2}\int\frac{d^3{\bf q}}{(2\pi)^3}{\rm e}^{-\left|{\bf
q}\right|^2 R^2/2}A({\bf q})\left(\left|{\bf
q}\right|^2+m^2\right)
\int\frac{d^3{\bf q}^{\prime}}{(2\pi)^3}{\rm e}^{-\left|{\bf
q}^{\prime}\right|^2 R^2/2}A({\bf q}^{\prime}) = \nonumber \\
& & \nonumber \\
& - & t^{2}\frac{\phi_0^2\:R_0^6}{\left(R_0^2+R^2\right)^3}\:
\left(m^2+\frac{3}{R_0^2+R^2}\right) \, .
\end{eqnarray}

Taking  $R_0\simeq R=\beta\xi$ and $\beta\gg 1$, we
see that the contribution to $\lgh\phi^2(t)\rg_V$ coming from
the gaussian configuration formed at $t=0$,
$\lgh\phi^2(t)\rg_V^{ho}$, decreases as $\sim t^2\phi_0^2 m^2$.

In the meanwhile, the contribution to
$\lgh\phi^2(t)\rg_V$ from the noise force,
$\lgh\phi^2(t)
\rg_V^{\gamma}$, starts to increase as
$t^3$:
\begin{eqnarray}
\lgh\phi^2(t)
\rg_V^{\gamma}& \simeq &\frac{2\:\eta\:T}{3}\:t^3\:
\int\:\frac{d^3{\bf q}}{(2\pi)^3}
\:{\rm e}^{-\left|{\bf
q}\right|^2 R^2}\nonumber\\
& & \nonumber \\
&=&\frac{1}{12\:\pi^{3/2}}\frac{\eta\:T}{R^3}\:t^3.
\end{eqnarray}
\vskip 5mm
\begin{flushleft}
{\it ii)} Long times:
\end{flushleft}
For $t\gg 1/\eta$, all the modes have relaxed to their
equilibrium values. The initial gaussian configuration
(\ref{initial}) has been rapidly thermalized with a relaxation
time $\tau_{\gamma}\simeq 1/\eta$ so that
$\lgh\phi^2(t)\rg_V^{ho}$ has quickly vanished.

The situation is again
similar to that for a brownian particle immersed in a thermal
environment at $t=0$ with initial data
$\vec{X}_0,\: \vec{v}_0$:
the contribution to $\langle\vec{X}^2(t)\rangle$ due to
the initial conditions is very rapidly
dissipated away and taken over by the
contribution from the noise.

As a
consequence, at long times
$\lgh\phi^2(t)\rg_V^{\gamma}$ provides the
only contribution to
$\lgh\phi^2(t)\rg_V$ which tends asymptotically towards the
{\it time-independent} dynamical
value
\begin{equation}
\label{equilibrium}
\lgh\phi^2\rg_V^{{\rm
dyn}}=\int\:\frac{d^3{\bf q}}{(2\pi)^3}
\:{\rm e}^{-\left|{\bf
q}\right|^2 R^2}\:\frac{T}{\left|{\bf
q}\right|^2+m^2}.
\end{equation}
Note that the dependence from the the viscosity coefficient $\eta$ in
the above expression is absent as implicit in the
fluctuation-dissipation theorem: the system after a sufficiently
long time relaxes towards its equilibrium state which is completely
independent
from $\eta$.

It is readily verified that the above expression coincides with the
{\it equilibrium} value of the
two-point connected Green's function, coarse-grained on a volume $V$,
calculated in ref. \cite{hr}. This is not surprising.
We already know that dynamics is expected to bring the
thermal fluctuations represented by the subcritical bubbles
into equilibrium with the surrounding thermal
bath at long times.

The result expressed by eq. (\ref{equilibrium})
shows explicitly that the  Langevin equation correctly describes
the long-time behaviour of
$\lgh\phi^2(t)\rg_V$ reproducing its equilibrium value if the
fluctuation-dissipation condition is satisfied, {\it i.e}
\begin{equation}
{\rm lim}_{t\rightarrow \infty}\:
\lgh\phi^2(t)\rg_V\equiv
\lgh\phi^2\rg_V^{{\rm
dyn}}=\lgh\phi^2\rg_V^{{\rm
EQ}}.
\label{fd}
\end{equation}
Note that eq.
(\ref{equilibrium}) still holds also in the case
$\eta>2\:m$. Namely, in the overdamped limit (large times
or strong coupling to the environment), the description
simplifies considerably, as the
canonical momentum associated to $\p$ becomes a
variable much faster than the field itself and can be
considered fully thermalized
in the adiabatic approximation.

The above statement amounts to drop the inertia
term $\ddot{\phi}({\bf x},t)$ in eq. (\ref{master}), leading
to the equation
\begin{equation}
\label{overdamped}
\eta\dot{\tilde{\phi}}\left({\bf q},t\right)+
\left(\left|{\bf
q}\right|^2+m^2\right)\ptilde=\tilde{\gamma}({\bf q},t),
\end{equation}
which reproduces the large time behaviour expressed by
eq. (\ref{equilibrium}).

Assuming $R=\beta/m$ and noting that the gaussian function in eq.
(\ref{equilibrium}) is substantially different from zero
only for $\left|{\bf q}\right|^2\lsim m^2/\beta^2$,
we get
\begin{equation}
\label{result}
\lgh\phi^2\rg_V^{{\rm
dyn}}\simeq \frac{m\:T}{2\pi^2}\left[\frac{1}{\beta}-{\rm tan^{-1}}\left(
\frac{1}{\beta}\right)\right].
\end{equation}
The spreading of the field over a correlation volume $V_\xi=(4\pi/3)\xi^3$
is then
\begin{equation}
\label{corr}
\lgh\phi^2\rg_{V_\xi}^{{\rm dyn}}\simeq 10^{-2}\:m\:T.
\end{equation}
For volumes $V$ substantially larger than the correlation
volume, $\lgh\phi^2\rg_V^{{\rm
dyn}}\simeq (m\:T/3\:\pi^2\:\beta^3)\ll m\:T$.
The coefficient $1/2\pi^2$ plays a crucial role in reducing the field
spreading, as already pointed out in refs. \cite{dine,bet,hr}.

One could
expect from these results that subcritical fluctuations
will not play a significant role in the electroweak phase transition.
Nevertheless, all the considerations made in this Section are limited to
a linear system. In the next Section we wish to extend them, through
a self-consistent Hartree approximation, to a
nonlinear system described by a $\lambda\phi^4$ potential as a further
step torwards the study of thermal fluctuations at the onset of
electroweak phase transition.
\vskip 1cm
\leftline{\large\bf 3. Self-interacting field: Hartree approximation}
\vskip 5mm
The above description of the evolution of
spherically-symmetric initial configurations
in the presence of a thermal bath is based on the linearity
of the Langevin equation.

Let us now consider the issue
of thermal fluctuations around the homogeneous configuration $\phi=0$
for a nonlinear system with a potential
\begin{equation}
\label{nonlinear}
V(\phi)=\frac{m^2}{2}\phi^2+\frac{\lambda}{4}\phi^4,
\end{equation}
where $m^2$ is assumed to be positive so that $V(\phi)$ posseses
a unique minimum at $\phi=0$.

In order to put the theory in a form suitable for the application
of the formalism developed in Section 2, we first resort to the Hartree
approximation. Taking the expectations over the free Gaussian measure
with their proper haffnian combinatoric factor, the Hartree prescription
yields the following substitution:
\begin{equation}
\label{sub}
\phi^4({\bf x},t)\rightarrow 6\lgh\phi^2\rg_V^{{\rm
dyn}}\:\phi^2({\bf x},t),
\end{equation}
where $\lgh\phi^2\rg_V^{{\rm
dyn}}$ is the asymptotic, time-independent
value of the two-point connected Green's function,
to be determined self-consistently by
equating its expression to the equilibrium
value $\lgh\phi^2\rg_V^{{\rm EQ}}$ derived from the dynamics.

In the Hartree theory the potential
$V(\phi)$ is then substituted by an {\it effective}
linear potential
\begin{equation}
V_{{\rm eff}}=\frac{m_{H}^2}{2}\phi^2,
\end{equation}
where the Hartree effective mass is now given by
\begin{equation}
\label{harmass}
m_{H}^2=m^2+3\lambda\lgh\phi^2\rg_V^{{\rm
dyn}}.
\end{equation}
The fact that $m_{H}^2$ is time-independent allows us to solve
the eq. (\ref{lanlin}) with the substitution $m^2\rightarrow m_{H}^2$
and to
make use
of the results given in Subsection 2.2. The contribution to
$\lgh\phi^2(t)\rg_V$ from the initial configuration
is expected to be dissipated away within a typical relaxation time
$\tau_{\gamma}$, whereas the contribution from the stochastic force
increases with time reaching for $t\gg 1/\eta$
the asymptotic value $\lgh\phi^2\rg_V^{{\rm
dyn}}$.

{}From eq. (\ref{equilibrium}) one can easily read off
the self-consistency condition as an intersection equation:
\begin{eqnarray}
\label{self}
\lgh\phi^2\rg_V^{{\rm
dyn}}&=&f\left(\lambda,\lgh\phi^2\rg_V^{{\rm
dyn}}\right) \equiv
\int\:\frac{d^3{\bf q}}{(2\pi)^3}
\:{\rm e}^{-\left|{\bf
q}\right|^2 R^2}\:\frac{T}{\left|{\bf
q}\right|^2+m_H^2}\nonumber\\
&& \nonumber \\
&=&\int\:\frac{d^3{\bf q}}{(2\pi)^3}
\:{\rm e}^{-\left|{\bf
q}\right|^2 R^2}\:\frac{T}{\left|{\bf
q}\right|^2+m^2+3\lambda\lgh\phi^2\rg_V^{{\rm
dyn}}}.
\end{eqnarray}
An upper bound for $\lgh\phi^2\rg_V^{{\rm
dyn}}$ then is trivially found observing that
\begin{equation}
f\left(\lambda,\lgh\phi^2\rg_V^{{\rm
dyn}}\right)\leq f\left(0,\lgh\phi^2\rg_V^{{\rm
dyn}}\right),
\end{equation}
which, taking into account the results of the previous Section, assures
that
\begin{equation}
\lgh\phi^2\rg_V^{{\rm
dyn}}\lsim 10^{-2}\:m\:T.
\end{equation}

This inequality is not surprising: being $m^2>0$, a positive value of $\lambda$
has the effect to increase the effective mass around $\phi=0$ making
fluctuations more difficult than in the case $\lambda=0$.

Some comments are in order here.
First of all, one can expect the
Hartree approximation to be reliable in the limit of small
coupling $\lambda$, {\it i.e.} when nonlinear effects can be
neglected in first approximation. In this respect, the Hartree
approximation amounts to consider the scalar
field $\phi$ to be the modulus of a vector field
$\vec{\varphi}$ with a large number $N$ of components
and self-coupling $(\lambda/N)\left|\vec{
\varphi}\right|^4$ \cite{ma}.

We have considered a nonlinear potential
(\ref{nonlinear}), that does not depend
upon the volume $V$ of the region where
one takes the space average of the
variance at long times.

This is strictly correct
\cite{hr}
only for regions with typical size $R\gsim \xi$ which are the ones we are
interested in. For  $R\lsim \xi$, one should modify expression
(\ref{nonlinear}) to properly take into account the finite size
effects.

In the next Section we finally address the role of thermal fluctuations
in the physically interesting case of the electroweak phase transition and
investigate whether the latter is hindered by the presence and the
equilibrium dynamics of the subcritical bubbles.

\vskip 1cm
\leftline{\large\bf 4. Fluctuations at the electroweak phase
transition}
\vskip 0.4cm
\leftline{\bf 4.1 Small supercooling limit and thin wall approximation}
\vskip 5mm
First order phase transition and critical
bubble dynamics in the Standard Model
have lately been studied in much detail, and it has become increasingly clear
\cite{clear} that for Higgs masses
considerably heavier than 60 GeV, the electroweak phase transition
is only of weakly first order.

For a Higgs mass
$M_H > 100$ GeV, both perturbative and lattice
calculations confront
technical problems.
However, it is conceivable that for such large
Higgs masses the electroweak phase transition is close to a second order
and does not proceed by critical bubble formation.

In this Section we consider a
phenomenological Higgs potential for the order parameter $\phi$ suitable
for a simple description of a first order phase transition:
\be{potential}
V(\phi ) = \frac 12 m^2(T)\phi^2 - \frac 13 \alpha T \phi^3 + \frac 14
\lambda\phi^4,
\ee
where we have not determined the parameters perturbatively but fit them, when
needed, according to a recent two-loop
determination of the gauge-invariant
effective potential \cite{lattice}.

Most of the dynamical properties of the electroweak
phase transition associated with the potential \eq{potential},
such as the smallness of the latent heat, the bubble
nucleation rate and the size of critical bubbles, have been discussed
in \cite{kari}. For the purposes of the present paper it suffices to
recall only some of the results.

Assuming that there is
only little supercooling, as seems to be the case for the electroweak
phase transition, the bounce action can be written as
\be{bounce}
S/T = {\alpha\over \lambda^{3/2}} {2^{9/2}\pi\over 3^5}
{\bar\lambda^{3/2}\over
(\bar\lambda - 1)^2}~,
\ee
where $\bar\lambda (T)=9\lambda m^2(T)/(2\alpha^2T^2)$.
The cosmological transition temperature is determined
from the relation that the Hubble rate equals the transition rate $\propto
e^{-S/T}$,
yielding $S/T_f \simeq {\rm ln} (M_{Pl}^4/T_f^4) \simeq 150$,
where $T_f$ is the transition temperature. Thus we obtain from \eq{bounce}
\be{lambdabar}
\bar \lambda(T_f) \simeq 1 - 0.0442{\alpha^{1/2}\over\lambda^{3/4}}\equiv
1-\delta.
\ee
On the other hand, small supercooling implies that $1-\bar\lambda=\delta\ll 1$,
i.e. $\alpha \ll 500 \lambda^{3/2}$. Solving for $\bar\lambda$ in \eq{bounce}
yields
the transition temperature $T_f$. One finds
\be{mass}
m^2(T_f) = {2 \alpha^2\over 9\lambda}\:\bar\lambda(T_f) \: T_f^2~.
\ee
The extrema of the potential are given by
\be{extrema}
\phi_\pm (T) = {\alpha T\over 2\lambda}(1 \pm\sqrt{1 - 8\bar\lambda/9}).
\ee
Expanding the potential at the broken minimum $\phi_{+}(T)$ we find
\be{epsilon}
-\epsilon\equiv V(\phi_{+},T_f)={1\over 6}m^2(T_f)\phi_{+}^2-
{1\over 12}\lambda\phi_{+}^4=-0.00218\:{\alpha^{9/2}\over
\lambda^{15/4}}
T_f^4+{\cal O}\left(\delta^2\right).
\ee
The height of the barrier is situated at $\phi_{-}\simeq
\phi_{+}/2$ with $V\left(\phi_{-},T_c\right)\equiv V_{{\rm
max}}=\alpha^4 T_c^4/(144\:\lambda^3)$, where $T_c$ is the temperature
at which $V(0)=V(\phi_{+})$, given by the condition $m(T_c)^2=(2\:
\alpha^2\: T_c^2/9\: \lambda)$. As $T_c\simeq T_f$ we may conclude that
the thin wall approximation is valid if $-\epsilon/V_{{\rm max}}=
0.314\:\alpha^{1/2}/\lambda^{3/4}\ll1$, or
$\alpha\ll 10\lambda^{3/2}$. Thus  the small supercooling
limit is clearly satisfied if the thin wall approximation is valid.

At $T=T_f$ the system can tunnel from the unbroken
phase $\phi=0$ to the broken phase
$\phi=\phi_{+}$ throught the formation of energetically favoured
critical bubbles with critical radius $R_c$.

To get the size of the critical bubble we still need the surface
tension. One easily finds
\be{surface}
\sigma=\int_{0}^{\infty}\:d\phi\sqrt{2\:V(T_c)}={2\:\sqrt{2}\:\alpha^3\over
91\:\lambda^{5/2}}\:T_c^3.
\ee
We define the critical bubble radius by extremizing the
bounce action. In the thin wall approximation the result is
\be{Rc}
R_c = 13.4\, {\lambda^{3/4}\over\alpha^{1/2}m(T_f)}.
\ee
Therefore $R_c$ is much
larger than the correlation length $\xi(T_f) = 1/m(T_f)$ at the transition
temperature, as it should. It is clear that fluctuations of the Higgs
scalar field $\phi$ with spatial correlations comparable to the critical
bubble size may be expected to be important for bubble nucleation.

\vskip 1cm
\leftline{\bf 4.2 Subcritical bubbles}
\vskip 5mm

Let us first make the general observation that
it is the actual transition temperature $T_f$ rather than the critical
temperature $T_c$
which is relevant in the study of subcritical bubbles.
This is true in the sense that if subcritical bubbles are not important
at $T_f$, they will not be at $T_c$ either.

As we shall show,
it actually turns out that subcritical bubbles are not important even
at  $T_f$. This justifies, in retrospect, our choice $T=T_f$ for performing
the calculations.

In the case of a
weak first order phase transition the critical bubble is typically well
described by a thin wall approximation, where the configuration is by
no means gaussian, but has a flat 'highland' (with $\phi$ determined
by the non-zero minimum of the potential) and a steep slope down to
$\phi=0$.

Therefore it seems natural that also a large subcritical
bubble should resemble the critical one: when its radius
$R$ increases,
the form of the subcritical bubble should deform smoothly so that, when
$R=R_c$, the bubble is a critical one.

The authors of ref. \cite{noi} took into account
this observation in their choice of the  ansatz $\phi=\phi(\phi_0,R)$
for the configuration describing the thermal fluctuations
around equilibrium.

Treating the amplitude $\phi_0$
(at the core) and the radius $R$ of such fluctuation as statistical
degrees of freedom, they found that
the average size $\langle R\rangle$ is much larger than the correlation
length $\xi(T_f)$ and that the most probable amplitude at the
core $\langle \phi_0^2\rangle^{1/2}$ is much smaller than $\phi_{-}(T_f)$,
the value of the field $\phi$ where the potential acquires its maximum.

It was then concluded \cite{noi} that subcritical bubbles, at least for
Higgs masses less than about 100 GeV, are not important during the
onset of the electroweak phase transition.

This conclusion was drawn also by taking into account the crucial role
played by thermalization: since the thermalization rate $\eta$ for small
amplitude scalar fluctuations and large spatial size was estimated in the
Standard Model
\cite{rate} to be of order of $10^{-2}\:T$ near the critical
temperature, {\it i.e.} much larger than the typical first order
transition time, small amplitude fluctuations with size larger than
about $1/\eta$ were considered to be absent from the mixture of
subcritical bubbles and not counted in the thermal averages.

However, we know that  the fluctuation-dissipation theorem describes the
common origin of dissipation and fluctuation: the
impact between the system and the surrounding particles of the
thermal bath is not limited
to a systematic friction force responsible for the dissipation
of fluctuations, but is also characterized by a random force responsible
for the fluctuations themselves. This stochastic force is fundamental,
as we have seen in Section 2, in
determining the dynamics of fluctuations with time and expecially
in the asymptotic regime.

{}From the considerations above, it should be clear that the thermal
averages $\langle R\rangle$ and $\langle \phi_0^2\rangle^{1/2}$ found
in ref. \cite{noi} must be respectively
regarded as the most probable size and
amplitude at the core of a subcritical bubble formed at $t=0$. Let us
then imagine that this "most probable" configuration appears at $t=0$.

To follow its dynamics, again we linearize the potential
applying the Hartree approximation:
\begin{equation}
\label{hartree}
\phi^3({\bf x},t)\rightarrow 3\lgh\phi^2\rg_V^{{\rm
dyn}}\:\phi({\bf x},t),\:\:\:\:\:\:\:\
\phi^4({\bf x},t)\rightarrow 6\lgh\phi^2\rg_V^{{\rm
dyn}}\:\phi^{2}({\bf x},t).
\end{equation}
The linearized Langevin equation now reads
\begin{equation}
\label{lanweak}
\ddot{\phi}({\bf x},t)-
\nabla^2\p+\eta\dot{\phi}({\bf x},t)+m_H^2\p -
\alpha\:T\:\lgh\phi^2\rg_V^{{\rm
dyn}}=\gamma\left({\bf x},t\right),
\end{equation}
where $m_H^2$ is the same in equation (\ref{harmass}).

Since we are assuming the electroweak phase transition to be first order, which
is expected to occur for for light Higgs masses or
equivalently $\lambda\ll 1$, and since
thermal fluctuations are expected to be important for a very weak first
order phase transition, {\it i.e.} for
$\alpha\ll 1$,  we expect the Hartree approximation for the quartic potential
(\ref{potential}) to be reliable since the coefficients of the nonlinear
terms in the potential are very small.

The Hartree approximation does
not allow to describe the nonlinear effects present in the theory.
For instance, Copeland {\it et al.} \cite{comment}
have recently pointed out that nonlinear scalar
field theories can incorporate
well localized, time-dependent, configurations, called
oscillons, which, although unstable, can be extremely long-lived,
$\tau_{sh}\simeq (10^3- 10^4)\:m^{-1}$. However, such configurations
can develop resonances only if they form with an initial
amplitude $\phi_0$ at the core above the inflection
point $\phi_{-}$, which seems to be not very probable \cite{noi}.
Moreover, oscillons were studied in absence of any coupling to a thermal
bath whose presence is expected to reduce considerably their lifetime.

Equation (\ref{lanweak}) is far from being the true Langevin equation
describing a thermal fluctuation in the hot electroweak theory. Indeed,
even in the simpler $\lambda\phi^4$ theory, the noise is colored and the
coupling to the thermal bath is multiplicative.

Also, the viscosity
coefficient $\eta$ is not expected to be a constant, but to depend upon
the typical momentum scale of the thermal fluctuation, {\it i.e.}
the inverse of its radius, and its amplitude. Up to now the only
estimate of $\eta$ for the Standard Model is  that given in ref.
\cite{rate} and is valid only for small amplitude $\phi_0$ and large size
$R$ fluctuations, $R\gsim 1/\eta\simeq 10^2/T$.

Fortunately enough, this
is just the range of $\phi_0$ and $R$ provided in ref. \cite{noi} for the
initial conditions that we have to specify for solving
the linearized Langevin equation; we therefore expect
that assuming a constant $\eta$ in momentum
space provides a reliable approximation.

Again, physical results
at long times, or at equilibrium, are
obviously expected to be independent from
the viscosity coefficient $\eta$ and from the nature of the thermal
noise, even if relaxation times can change significantly.

Equation (\ref{lanweak}) is easily reduced in a form equivalent
to the one holding in the free field case by absorbing the constant
non-homogenous term through the following linear redefinition
of the field variable:
\begin{eqnarray}
\label{pos}
\p&=&\varphi({\bf x},t)+\kappa,\nonumber\\
&& \nonumber \\
\kappa&=&\frac{\alpha\:T}{3\:\lambda+(m_H^2/\lgh\phi^2\rg_V^{{\rm
dyn}})},
\end{eqnarray}
where $\varphi({\bf x},t)$ solves the Langevin equation
(\ref{lanlin}) with $m^2\rightarrow m_H^2$. Note that $\kappa$ in the
limit of very large
$\lgh\phi^2\rg_V^{{\rm
dyn}}$ approaches $\phi^2_{-}$.

For definiteness, we fitted our phenomenological potential
\eq{potential} to the two loop result for the effective
potential calculated in \cite{lattice} for the Higgs mass $M_H=70$ GeV.
This yields $\alpha\simeq 0.048$ and $\lambda\simeq 0.061$.
One readily verifies that we are indeed safely in the thin wall limit.
With these parameter values ({\it in units of $R_c$})
\begin{eqnarray}
T_f&=&85.70,\nonumber\\
m^2(T_f)&=&56.78\nonumber\\
\eta&\simeq& 10^{-2}\:T_f\simeq 0.875\nonumber\\
\phi_{-}(T_f) & \simeq & 24.
\end{eqnarray}
Assuming to be again in the situation $\eta<2\:m_H$ (we shall verify
{\it a posteriori} that this is indeed the case),
we easily
read off the solution of the linearized Langevin eq. (\ref{lanweak}):
\begin{eqnarray}
\label{solution}
\p&=&\varphi({\bf x},t)+\kappa\:\left(
1-{\rm e}^{-\eta
t/2}\cos\frac{\sqrt{\eta^2-4\:m_H^2}}{2}t\right.\nonumber\\
&& \nonumber \\
&-&\left.
{\rm e}^{-\eta t/2}\:\frac{\eta}{\sqrt{\eta^2-4\:m_H^2}}\:\sin
\frac{\sqrt{\eta^2-4\:m_H^2}}{2}t\right),
\end{eqnarray}
where $\varphi({\bf x},t)$ does have the same time evolution described in
Section 2.

Dynamics evolves as follows. At $t=0$ a subcritical bubble is
formed with initial amplitude at the core $\phi_0$ and radius $R$ given in ref.
\cite{noi}. For $t\ll 1/\eta$, $\lgh\phi^2(t)\rg_V$ none of the modes of
the initial configuration have yet relaxed to zero, even if they start
to decrease as $t^{2}$. In the meanwhile, besides the noise term
$\lgh\phi^2(t)\rg_V^{\gamma}$ which increases as $t^3$, a new term begins
to contribute to $\lgh\phi^2\rg_V$, {\it i.e.} the term proportional to
$\kappa^2$ increasing as $t^2$. For $t\gg 1/\eta$, all the modes have
relaxed towards their equilibrium value. The initial
configuration has disappeared, and $\lgh\phi^2\rg_V^{{\rm
dyn}}$ is the sum of the noise term contribution and the $\kappa^2$-term
contribution, so that the self-consistency equation for $\lgh\phi^2\rg_V^{{\rm
dyn}}$ reads
\begin{eqnarray}
\label{f}
\lgh\phi^2\rg_V^{{\rm
dyn}}&=&f\left(\lgh\phi^2\rg_V^{{\rm
dyn}}\right)\nonumber\\
&& \nonumber \\
&=&\kappa^2+\int\:\frac{d^3{\bf q}}{(2\pi)^3}
\:{\rm e}^{-\left|{\bf
q}\right|^2 R^2}\:\frac{T}{\left|{\bf
q}\right|^2+m^2+3\lambda\lgh\phi^2\rg_V^{{\rm
dyn}}}.
\end{eqnarray}
In Fig. 1 we present the behaviour of the function
$f\left(\lgh\phi^2\rg_V^{{\rm
dyn}}\right)$ (dashed line) in units of $1/R_c$ for $\lambda=0.061$,
$\alpha=0.048$
and $R=\xi(T_f)$, {\it i.e.} the variance is calculated
in a typical correlation volume. Note that in the limit of very large
$\lgh\phi^2\rg_V^{{\rm
dyn}}$, $f$ tends toward $\phi_{-}^2(T_f)\simeq (24/R_c)^2$.

Fig. 2 gives the same
function for the two
cases $R=\xi(T_f)$ and $R=2\:\xi(T_f)$. The points where the
dashed lines intersect the solid line, representing the left hand
side of eq. (\ref{f}), denote the values  of $\lgh\phi^2\rg_V^{{\rm
dyn}}$ which satisfy the self-consistency equation in the two different
cases: when testing the fluctuation in a correaltion volume $V_\xi=
(4\pi/3)\xi^3$, we have
\begin{equation}
\sqrt{\lgh\phi^2\rg_{V_{\xi}}^{{\rm
dyn}}}\simeq (2.7/R_c)\ll \phi_{-}(T_f)\simeq (24/R_c);
\end{equation}
if we test the fluctuation in a volume
larger than $V_\xi$, the variance is even smaller,
\begin{equation}
\sqrt{\lgh\phi^2\rg_{V}^{{\rm
dyn}}}\simeq (1.2/R_c)
\end{equation}
for $R=2\:\xi(T_f)$. Note also that
our starting assumption $\eta<2\:m_H$ is verified.

One could naively think that the extra energy due to a
nonvanishing variance would facilitate barrier penetration for the
formation of critical bubbles or possibly even invalidate our thin wall
approach.

We can verify that this is not the case by considering the
fluctuation energy about $\phi=0$. For $R=\xi(T_f)$ it is given
approximatly by (in units of $1/R_c^4$)
\begin{equation}
\label{energy}
E_{{\rm fl}}\simeq \frac{1}{2}m^2(T_f)\lgh\phi^2\rg_{V}^{{\rm
dyn}}\simeq 211\ll -\epsilon\simeq 5\times 10^3,
\end{equation}
where we have made use of the fact that $\phi_{+}(T_f)\simeq
(48/R_c)$. Thus, in terms of the barrier penetration, thermal
fluctuations represent only a minor correction and they do not have
remarkable effect on the thin wall approximation.

{}From these considerations and from the fact that
the variance of the thermal fluctuations
$\sqrt{\lgh\phi^2\rg_{V}^{{\rm
dyn}}}$
is clearly smaller than the
inflection point $\phi_{-}(T_f)$, we can conclude that subcritical
bubbles do not affect the nucleation of critical bubbles in an
appreciably way.

\vskip 1cm
\leftline{\large\bf 5. Conclusions and perspectives for future work}
\vskip 5mm
In this paper we have investigated the role played by subcritical
bubbles at the onset of the electroweak phase transition.

Treating the
the configuration $\p$ modelling such thermal fluctuation as a
stochastic variable, we have described its dynamics by a
phenomenological Langevin equation for different models.
This approach allows to properly
take into account both effects of the thermal bath on the system:
a systematic dissipative force, which tends to erase out any initial
subcritical configuration, and a random stochastic force responsible for
fluctuations.

Following the evolution with time of the variance
$\lgh\phi^2(t)\rg_V$ in a given volume $V$, we have shown that the
contribution to it from any initial subcritical configuration is
quickly damped away and, in the limit of long times,
$\lgh\phi^2(t)\rg_V$ approaches its equilibrium value provided by the
stochastic force and independent from the viscosity coefficient $\eta$,
in agreement with the fluctuation-dissipation theorem.

In the most interesting case of thermal fluctuations
during the electroweak phase transition, we have made use of a
self-consistent Hartree approximation expected to
give good results in the weak
coupling limit which we believe to hold for a weak first order
phase transition. A more correct way of proceeding might have
been to make use of the variance $\lgh\phi^2(t)\rg_V$ in the
substitution (\ref{hartree}) instead of $\lgh\phi^2\rg_{V}^{{\rm
dyn}}$. However, since the system reaches very quickly the equilibrium,
we are confident that our approximation is reliable.

We have shown that thermal fluctuations do not
affect the nucleation of critical bubbles at the tunneling temperature
$T_f$ and from this fact we conclude that electroweak baryogenesis
scenarios associated to a weak first order electroweak phase
transition remain a viable possibility to explain the primordial baryon
asymmetry in the Universe.

Very recently
Bettencourt \cite{bet} has  computed the probability {\it at
equilibrium} for the
fluctuations of the Standard Model Higgs field, averaged over a
correlation
volume, to exceed the inflection point $\phi_{-}$ at $T\simeq T_f$ and
for a physical Higgs mass
$M_H=70$ GeV, the same value we have used in our paper.

He estimated
such a probability to be very small, around 1.27\%, and to
decrease when the testing volume $V$ is increased. These estimates  and
behaviour with $V$ are in
complete agreement with our results for the asymptotic value of
$\lgh\phi^2(t)\rg_{V}$ whose evolution with time is also given here.

The phenomenological Langevin equation used in our paper
is not the {\it true} equation describing a thermal fluctuation in the
hot electroweak theory.

In this respect,
our work should be regarded as a first step towards a complete
description of the dynamics of thermal fluctuations at the
electroweak phase transition
by a more complicated Langevin equation which properly
describes the coupling of
the stochastic field $\p$ to the other degrees of freedom.

Nevertheless,  we expect our physical results to be reliable
at long times when the details on the nature of the thermal noise and
the viscosity coefficient $\eta$ become unessential. On the other hand,
relaxation time-scales crucially depend on the nature of the
stochastic force and the strength of dissipation and their complete
knowledge is needed to decide if degrees of freedom other than $\p$
are in equilibrium or not inside the subcritical bubble.

One may realize that this is a
crucial question by reminding that the effective
potential (\ref{potential}), used to describe the free
energy associated to the fluctuations,
is usually obtained at 1-loop integrating out all the degrees of freedom of the
theory other than $\p$, {\it i.e.} fermions and gauge bosons.

In
performing such a calculation, it is assumed that fermions and gauge
bosons do have {\it equilibrium} distributions with a $\p$ background
dependent mass. This is true only if their interaction times with the
background $\p$ are much smaller than the typical lifetime of the
subcritical bubble.

Since this condition seems not be satisfied, a full
non-equilibrium approach is needed. Such an approach
might  lead to unexpected results
as far as the issue of thermal fluctuations during the electroweak phase
transition is concerned \cite{iiro}.
\vskip 1cm
\centerline{\large\bf Acknowledgements}
\vskip 0.3cm
It is a pleasure to thank K. Enqvist and I. Vilja for carefully reading the
manuscript and for useful suggestions.

\newpage

\newpage
\centerline{\large\bf Figure Captions}
\vskip 0.3cm

{\bf Figure 1}: The plot of the function $f\left(\lgh\phi^2\rg_V^{{\rm
dyn}}\right)$ (dashed line) in units of $1/R_c^2$ for $\lambda=0.061$, $
\alpha=0.048$ and $R=\xi(T_f)$. The solid line represents the inflection
point $\phi_{-}^2(T_f)$.
\vskip 0.1cm

{\bf Figure 1}: The plot of the function $f\left(\lgh\phi^2\rg_V^{{\rm
dyn}}\right)$ (dashed lines) in units of
$1/R_c^2$ for the same values of $\lambda$ and $
\alpha$ given in Figure 1 and for $R=\xi(T_f)$ and $R=2\:\xi(T_f)$
respectively. The intersection points between the dashed lines and the
solid line represent the values of $\lgh\phi^2\rg_V^{{\rm
dyn}}$ satisfying the self-consistency equation in the two cases, see
the text.


\begin{thebibliography}{99}
\bibitem{ckn} For a review, see A.G. Cohen, D.B. Kaplan and A.E. Nelson,
Ann. Rev. Nucl. Part. Phys. {\bf 43}, 27 (1993) ; D.B. Kaplan,
contribution
to the 4$^{{\rm th}}$ International Conference on Physics Beyond the
Standard Model, Lake Tahoe, 13-16 December 1994, hepph 9503360.

\bibitem{sak} A.D. Sakharov, Pis'ma Zh. Eksp. Fiz. {\bf 5},
32 (1967); JETP Lett. {\bf 5}, 24 (1967).


\bibitem{gleiser} M. Gleiser, E.W. Kolb and R. Watkins, Nucl. Phys.
{\bf B364}, 411 (1991); M. Gleiser and E.W. Kolb, Phys. Rev. Lett. {\bf
69}, 1304 (1992); Phys. Rev. {\bf D48}, 1560 (1993);
M. Gleiser and R.O. Ramos, Phys. Lett. {\bf B300}, 271 (1993);
N. Tetradis, Z. Phys. {\bf C57}, 331 (1993).

\bibitem{dine} M. Dine, R. Leigh, P. Huet, A. Linde and D. Linde,
Phys. Rev. {\bf D46}, 550 (1992).

\bibitem{and} G. Anderson, Phys. Lett. {\bf B295},
32 (1992).


\bibitem{gel} G. Gelmini and M. Gleiser, Nucl. Phys. {\bf B419},
459 (1994).

\bibitem{lep} Particle Data Group, {\it Review of Particle Properties,}
Phys. Rev. {\bf D50}, 1173 (1994).

\bibitem{noi} K. Enqvist, A. Riotto and I. Vilja, HU-TFT/95-32 preprint,
hep-ph 9505341, submitted to Phys. Rev. {\bf D}.

\bibitem{bet} L.M.A. Bettencourt, Imperial/TP/94-95/38 preprint.

\bibitem{hr} M. Hindmarsh and R.J. Rivers, Nucl. Phys. {\bf B417},
 506 (1994); R.J. Rivers, {\it Fluctuations at Phase Transitions},
in the Proceedings of the {\it Nato Advanced Research Workshop on Electroweak
Physics and the Early Universe}, Sintra, Portugal, March 1994.

\bibitem{kardar} See, for instance, M. Kardar, {\it Lectures on
Dynamics of Interfaces} in {\it recent Advances in Statistical physics},
Istanbul Technical University, Istanbul, Turkey, July 15 to August 7, 1993.

\bibitem{kpz} M. Kardar, G. Parisi and Y.-C. Zhang, Phys. Rev. Lett.
{\bf 56}, 889 (1986).

\bibitem{hoh} P.C. Hohenberg and B.I. Halperin, Rev. Mod. Phys.
{\bf 49}, 435 (1977).


\bibitem{vari} M. Gleiser and R.O. Ramos, Phys. Rev. {\bf D50},
2441  (1994);
B.L. Hu, J.P. Paz and Y. Zhang, in {\it The Origin of Structure in
the Universe}, Ed. E. Gunzig and P. Nardone (Kluwer Acad. Publ. 1993);
D. Lee and D. Boyanowsky, Nucl. Phys. {\bf B406}, 631  (1993); S. Habib,
in {\it Stochastic Processes in Astrophysics}, Proc. Eigth Annual
Workshop in Nonlinear Astronomy (1993),

\bibitem{der} G.H. Derrick, J. Math. Phys. {\bf 5}, 1252 (1964) .


\bibitem{comment} See also
E.J. Copeland, M. Gleiser and H.-R. Muller, Fermilab-Pub-95/021-A
preprint.

\bibitem{ma} S.K. Ma, Rev. Mod. Phys. {\bf 45}, 589 (1973).

\bibitem{clear}K. Kajantie, K. Rummukainen and M. Shaposnikov, Nucl.
Phys. {\bf B407}, 27 (1993); K. Farakos, K. Kajantie, K. Rummukainen and
M.E. Shaposhnikov, Phys. Lett. {\bf B336}, 494 (1994); B. Bunk, E.M.
Ilgenfriz, J. Kripfganz and A. Schiller, Nucl. Phys. {\bf B403}, 453 (1993);
F. Csikor {\it et al.}, Phys. Lett. {\bf B334}, 405 (1994).

\bibitem{lattice} Z. Fodor and Hebecker, Nucl. Phys. {\bf B432}, 127 (1994).


\bibitem{kari} K.\ Enqvist, J.\ Ignatius, K.\ Kajantie and K.\ Rummukainen,
 Phys. Rev. {\bf D45}, 3415 (1992).

\bibitem{rate} P.\ Elmfors, K.\ Enqvist and I.\ Vilja, Nucl.\ Phys.\
{\bf B412}, 459 (1994).

\bibitem{iiro} I. Vilja and A. Riotto, in preparation.

\end{thebibliography}
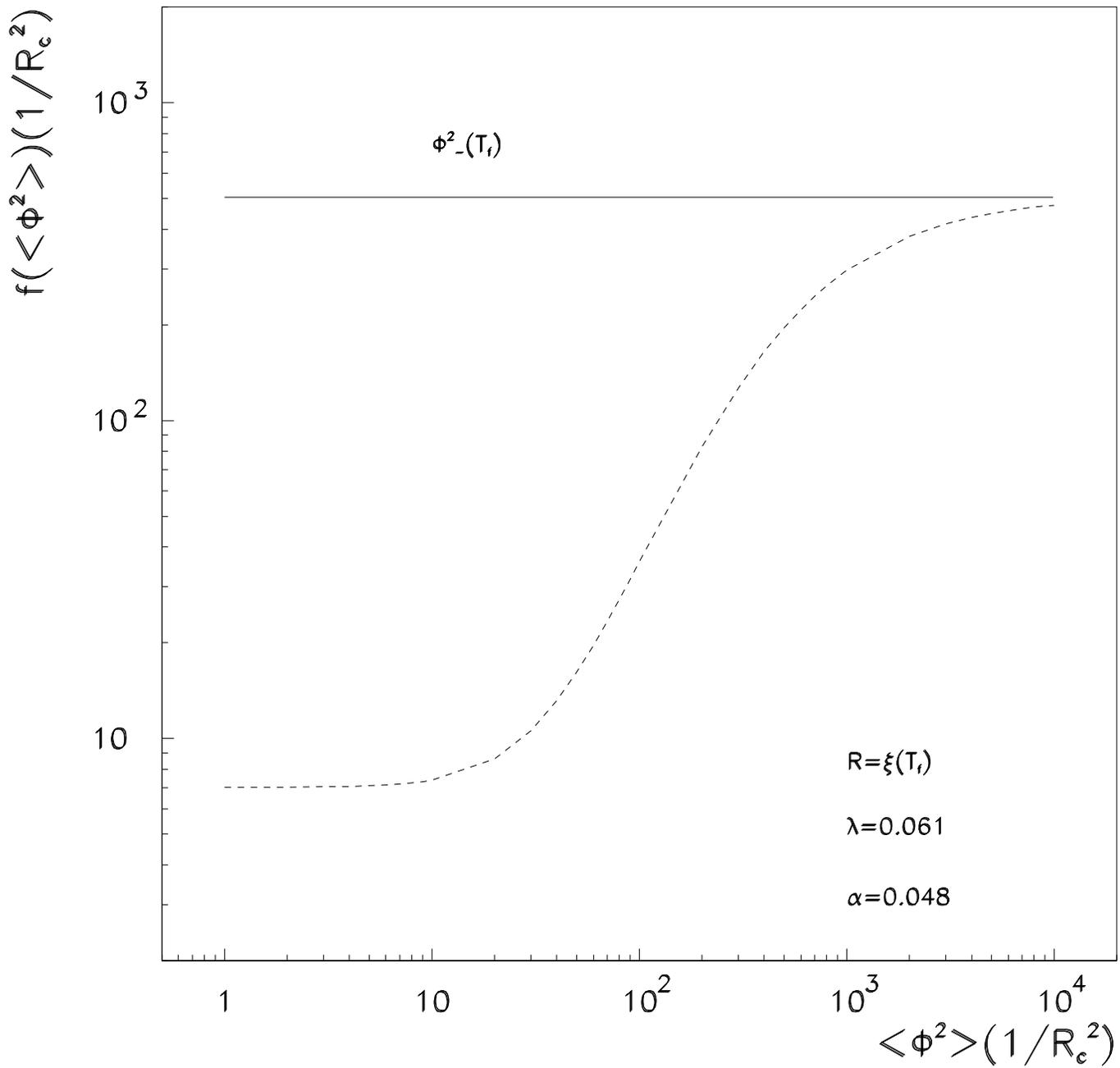
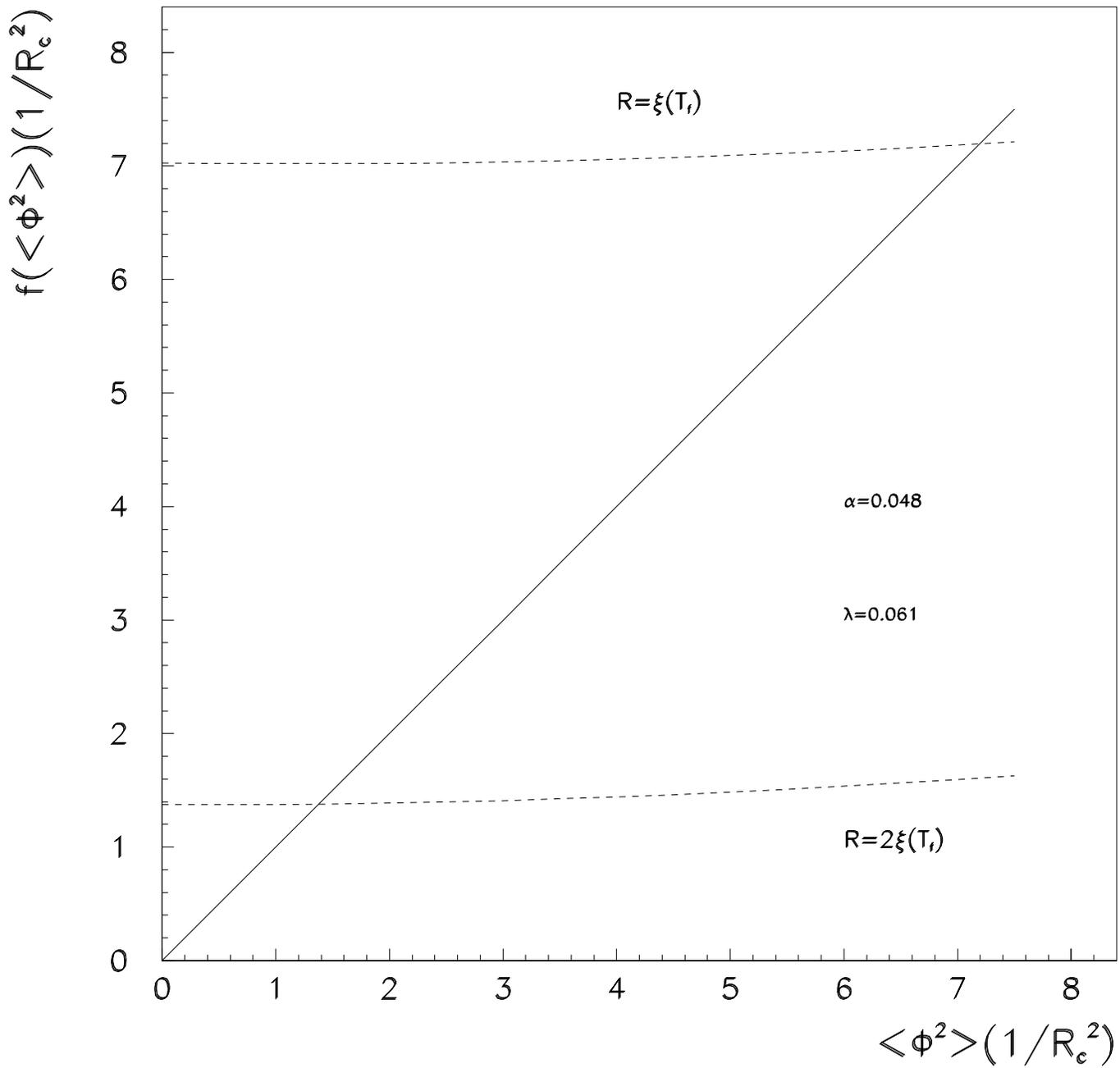
\end{document}